\documentclass[aps,prd,onecolumn,showkeys,showpacs,amssymb]{revtex4}
\usepackage{float}
\usepackage{amssymb}
\usepackage{graphicx}
\usepackage{amsmath}
\usepackage{hyperref}
\begin{document}

\title{Maximal neutron star mass and the resolution of hyperonization puzzle in modified gravity}

\author{Artyom V. Astashenok$^{1}$, Salvatore Capozziello$^{2,3,4}$, Sergei D. Odintsov$^{5,6,7}$}

\affiliation{$^{1}$I. Kant Baltic Federal University, Institute of Physics and Technology, Nevskogo st. 14, 236041 Kaliningrad, Russia,\\
$^2$Dipartimento di Fisica, Universita' di Napoli "Federico II and
\\$^3$INFN Sez. di Napoli, Compl. Univ. di Monte S. Angelo, Ed.
G., Via Cinthia,
9, I-80126, Napoli, Italy,\\
$^4$ Gran Sasso Science Institute (INFN),  Viale F. Crispi, 7, I-67100, L'Aquila, Italy\\
$^5$Instituci\`{o} Catalana de Recerca i Estudis Avan\c{c}ats (ICREA), Barcelona, Spain\\
$^6$Institut de Ciencies de l'Espai (IEEC-CSIC), Campus UAB, Torre C5-Par-2a pl, E-08193 Bellaterra, Barcelona, Spain\\
$^7$Tomsk State Pedagogical University (TSPU), Tomsk, Russia}

\begin{abstract}
The so-called ``{\it hyperonization puzzle}'' in the theory of neutron stars is
considered in the framework  of modified $f(R)$ gravity. We show that for
simple hyperon equations of state,  it is possible to obtain
the  maximal neutron star mass which satisfies the recent
observational  data for PSR J1614-2230, in  higher-derivative models
with power-law terms  as $f(R)=R+\gamma R^{2}+\beta R^3$. The soft
hyperon equation of state under consideration is usually treated as
non-realistic  in the standard General Relativity. The
numerical analysis of Mass-Radius relation for massive neutron
stars with hyperon equation of state in modified gravity turns out to be
consistent with observations. Thus, we show that the same modified
gravity can solve at once three problems: consistent description
of the maximal mass of neutron star, realistic Mass-Radius relation
and  account for hyperons in equation of state.
\end{abstract}


\maketitle

\section{Introduction}

The recent discovery of the pulsar PSR J1614-2230 \cite{Demorest}
has set  rigid constraints on various matter equations of state (EOS)
for  neutron stars at high densities. There are other
indications in favor of the existence of massive neutron stars:
$1.8M_{\odot}$ for Vela X-1 \cite{Rawls} and $2M_{\odot}$ for 4U
1822-371 \cite{Munoz}. In particular, this new limits on maximal
mass of neutron star excluded many EOS, including hyperons and/or
quarks EOS. According to the experimental data \cite{Nagae} and
realistic models for strong interactions, the appearance of exotic
particles occurs at densities $5-8\times 10^{14}$ g/cm$^{3}$.
However, the hyperonisation softens EOS and the maximal allowable
mass is reduced considerably \cite{Glendenning, Glendenning-2, Schaffner, Vidana,Schulze}.
The neutron stars with $M=2M_\odot$ cannot be
obtained in the framework of Thomas-Fermi model for non-uniform matter
\cite{Shen} with hyperon inclusion \cite{Ishizuka,Shen-2}
or a quark-hadron phase transition \cite{Nakazato}.

The solution of such a ``{\it hyperonization puzzle}'' can be searched, in
principle, by constructing the hyperon EOS giving  the
maximal mass of neutron star  around $2M_{\odot}$. The
required stiffness of the EOS can be achieved in relativistic mean
field theory (RMF) with hyperon-vector coupling larger than it
follows from $SU(6)$ symmetry models \cite{Hofmann, Rikovska}. A
model with chiral quark-meson coupling with
$M_{max}=1.95M_{\odot}$ has been recently considered \cite{Miyatsu}. The
quartic vector-meson terms in the Lagrangian also lead to the
stiffening of EOS and large neutron star mass \cite{Bednarek}. The
radius measurements of neutron stars could give more information
about EOS for dense matter. Unfortunately, one has no such
measurements for any neutron stars with a precise mass
determination. Nevertheless, there are some astrophysical
observations that could lead to the extraction of neutron star
radii \cite{Lattimer}.

As shown in \cite{Ozel}, based on data for radii and masses of
three neutron stars (in EXO 1745-248 \cite{Ozel-2}, in 4U 1608-52
\cite{Guver} and in 4U 1820-30 \cite{Guver-2}), the EOS with only
nucleonic degrees (such as AP4, MP1) are too stiff at higher
density.  A softer EOS describes these data with better precision.
In Fig. 1 these data and the theoretical $M-R$ relation for some
hyperon EOS are shown. We give this relation for simple model with
hyperons, proposed by Glendenning and Moszkowski (GM-model with
three parameterizations, GM1-3, see\cite{Glendenning,
Glendenning-2} for details). For illustration we also include the
EOS with quarks (pcnphq). The following feature is obvious:
although soft hyperon EOS predict the maximal value of mass $<2
M_{\odot}$, these EOS are more compatible with data by
\cite{Ozel-2, Guver, Guver-2}.

\begin{figure}
  \includegraphics[scale=1.1]{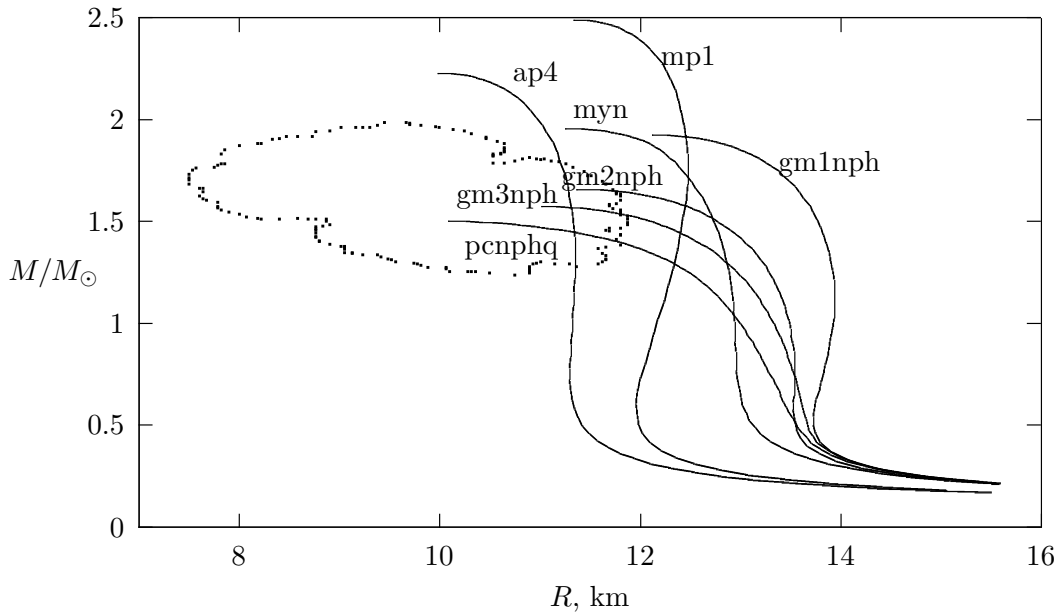}\\
  \caption{The mass-radius diagram for neutron stars coming from
some hyperon EOS (gm1nph, gm2nph, gm3nph, myn, pcnphq) and
nucleonic EOS (ap4 and mp1) for comparison. The abbreviation
``myn'' means EOS with chiral quark-meson coupling proposed
recently by Miyatsu, Yamamuro and Nakazato \cite{Miyatsu} (this
EOS is available in table form). For another EOS, we use
analytical representations by \cite{Eksi}. The constraints derived
from observations of three neutron stars by \cite{Ozel} are
represented by the dotted contour.}
\end{figure}

In Ref.\cite{Weissenborn}, it is shown that, in the case of
hyperonic matter with three exchange meson fields, the maximal mass
is achieved only for low values of the effective nucleon mass. The
addition of strange meson $\psi$ allows to increase the maximal
value of effective mass. Therefore, for the explanation of the new
maximal limit of neutron star mass, one needs to complicate the
simple '$\rho\omega\sigma$'-model. In fact, this complication
leads to the  stiff hyperon EOS which are close to pure nucleonic
EOS (such as MP1) and we have contradiction with the data
in \cite{Ozel-2, Guver,Guver-2}.

One note also that studying of longer X-ray bursts
\cite{Suleimanov}, \cite{Hambaryan} gives the relatively large
radii ($R>14$ km) for masses $\sim 1-1.3M_{\odot}$. This can say
in favor to models with simple hyperon EOS which predict radii
$R>\sim13.5$ km for $1-1.3 M_{\odot}$ stars. However these models
don't give the required upper mass limit.

One can assume that such contradictions can be considered as a
further indication in favor of the necessity to re-examine
gravity theory at the early/late universe or in strong field
regimes.

The initial motivation for this approach has been pursued starting
from the observed
 accelerated expansion of the early/late universe. This fact has been
confirmed by observations data. First of all, type Ia supernovae
point out an accelerated expansion which cannot be obtained by
standard perfect fluid matter as the source for the cosmological
Friedman-Robertson-Walker equations \cite{Perlmutter, Riess1, Riess2}. Second, one can
 mention the observations  of microwave background radiation (CMBR)
anisotropy \cite{Spergel}, of cosmic shear through gravitational
weak leasing surveys \cite{Schmidt} and, finally, data coming from Lyman
alpha forest absorption lines \cite{McDonald}. To explain the
universe acceleration within General Relativity (GR), one needs to
postulate the existence of some cosmic fluid with negative pressure
(dark energy). In the framework of $\Lambda $CDM model, dark energy is
nothing else but the Einstein Cosmological Constant and its density
is about 70\% of the global energy budget of the universe. The
remaining 30\%, clustered in galaxies and clusters of galaxies,
should be constituted only for about 4\% by baryons and for the
rest by cold dark matter (CDM) the nature of which is, up to now, unclear.

Despite of the simplicity and the good agreement with
observational data, the $\Lambda$CDM model has some fundamental
problems at theoretical level. For example, one needs to  explain
the difference of 120 orders of magnitude between its observed
value at cosmological level and the one predicted by
 quantum field theory/gravity \cite{Weinberg}.

From another viewpoint, the accelerated expansion of the universe (without
dark components) maybe naturally explained by modification of
gravity at the very early and very late universe. Indeed, modified
gravity may provide viable gravitational candidate for dark energy
(see refs. \cite{Capozziello1, Odintsov1, Turner} as well as for
unification of dark energy and early-time
inflation\cite{Odintsov1} (for recent review of modified gravity,
see \cite{Odintsov-3, Capozziello_book, Capozziello4, Cruz}). For
instance,it has been  shown that such theories give models which
are able to reproduce the Hubble diagram derived from SNela
observations \cite{Capozziello4,Demianski} and the anisotropies of
CMBR \cite{Perrotta, Hwang}.

Addressing the problem of  exotic relativistic stars in modified gravity, in comparison
with GR, could represent a testbed  for modified gravity. For example,
some models of $f(R)$ gravity do not allow the existence of stable
star configurations \cite{Briscese, Abdalla, Bamba,
Kobayashi-Maeda, Nojiri5,Dolgov,Sami,Ba,Bamba} and thus are
considered  unrealistic. However the existence of stable star
configurations can be achieved  in certain cases due to the
so-called {\it Chameleon Mechanism} \cite{Justin, Upadhye-Hu} or
may depend on the  chosen EOS.

In this paper, we present the models of neutron star for simple
hyperon EOS with maximal mass $\sim 2M_{\odot}$ in the framework of
analytic  $f(R)$ models. We show that it is
possible to address simultaneously the maximal value of neutron
star mass as well as fit the data by \cite{Ozel} assuming a
 hyperon EOS for dense matter. The  paper is organized as
follows. In Section II, we investigate the field equations for
$f(R)$ gravity and the modified Tolman--Oppenheimer--Volkoff (TOV)
equations. Then neutron star models with hyperon EOS in power-law
modified gravity are considered. Mass-Radius diagram is derived
and compared with the one of GR. The possibility
to get maximal mass for neutron stars and consistent Mass-Radius
relation for hyperon EOS within modified gravity is established.
Conclusions and outlook are given in Section III.

\section{Modified TOV equations in $f(R)$ gravity}

The general action for $f(R)$ gravity is given by
\begin{equation}\label{action}
S=\frac{c^4}{16\pi G}\int d^4x \sqrt{-g}f(R) + S_{{\rm
matter}}\quad.
\end{equation}
Here $g$ is the determinant of the metric $g_{\mu\nu}$ and $S_{\rm
matter}$ is the action of the standard perfect fluid matter. The
variation of (\ref{action}) with  respect to  $g_{\mu\nu}$ gives
the field equations.  The function $f(R)$ can be written as
\begin{equation}\label{fR}
    f(R)=R+\alpha h(R)\,,
\end{equation}
putting in evidence the extra contributions with respect to GR.
The field equations are
\begin{equation}\label{field}
(1+\alpha h_{R})G_{\mu \nu }-\frac{1}{2}\alpha(h-h_{R}R)g_{\mu \nu
}-\alpha (\nabla _{\mu }\nabla _{\nu }-g_{\mu \nu }\Box
)h_{R}=8\pi G T_{\mu \nu }/c^{4}.
\end{equation}
Here $G_{\mu\nu}=R_{\mu\nu}-\frac12Rg_{\mu\nu}$ is the Einstein
tensor and ${\displaystyle h_R=\frac{dh}{dR}}$.

For the star configurations,  one can assume a spherically symmetric
metric with two independent functions of radial coordinate, that
is:
\begin{equation}\label{metric}
    ds^2= -e^{2\phi}c^2 dt^2 +e^{2\lambda}dr^2 +r^2 (d\theta^2
    +\sin^2\theta d\phi^2).
\end{equation}

Then the following change of variable can result convenient
\cite{Stephani,Cooney}
\begin{equation}\label{mass}
    e^{-2\lambda}=1-\frac{2G M}{c^2 r}.
\end{equation}
For the exterior solution, we assume a Schwarzschild solution and
therefore the value of variable $M$ on the star surface is nothing
else but  the  gravitational mass.
 For a perfect fluid,  the energy-momentum tensor is $T_{\mu\nu}=\mbox{diag}(e^{2\phi}\rho
c^{2}, e^{2\lambda}P, r^2P, r^{2}\sin^{2}\theta P)$,  where $\rho$ is the
matter density and $P$ is the pressure. The components of the
field equations become
\begin{eqnarray}
  -8\pi G \rho/c^2 &=& -r^{-2} +e^{-2\lambda}(1-2r\lambda')r^{-2}
                +\alpha h_R\left[-r^{-2} +e^{-2\lambda}(1-2r\lambda')r^{-2}\right] \nonumber \\
             && -\frac12\alpha(h-h_{R}R) +e^{-2\lambda}\alpha[h_R'r^{-1}(2-r\lambda')+h_R''] \label{f-tt},\\
  8\pi G P/c^4 &=& -r^{-2} +e^{-2\lambda}(1+2r\phi')r^{-2}
                +\alpha h_R\left[-r^{-2} +e^{-2\lambda}(1+2r\phi')r^{-2}\right] \nonumber \\
             && -\frac12\alpha(h-h_{R}R) +e^{-2\lambda}\alpha h_R'r^{-1}(2+r\phi'), \label{f-rr}
\end{eqnarray}
where $'\equiv d/dr$.

The combination of the conservation law equation
 with Eq.(\ref{f-rr}) allows to obtain the second TOV
equation.

Finally, modified TOV equations take the following convenient
form \cite{Astashenok} (see, also \cite{capo})
\begin{equation}\label{TOV-1}
\left(1+\alpha  h_{{R}}+\frac{1}{2}\alpha h'_{{R}}
r\right)\frac{dm}{dr}=4\pi{\rho}r^{2}-\frac{1}{4}\alpha r^2
\left[h-h_{{R}}{R}-2\left(1-\frac{2m}{r}\right)\left(\frac{2h'_{{R}}}{r}+h''_{{R}}\right)\right],
\end{equation}
\begin{equation}\label{TOV-2}
8\pi p=-2\left(1+\alpha
h_{{R}}\right)\frac{m}{r^{3}}-\left(1-\frac{2m}{r}\right)\left[\frac{2}{r}(1+\alpha
h_{{R}})+\alpha r_{g}^{2}
h'_{{R}}\right]({\rho}+p)^{-1}\frac{dp}{dr}
\end{equation}
$$
-\frac{1}{2}\alpha
\left[h-h_{{R}}{R}-4\left(1-\frac{2m}{r}\right)\frac{h'_{{R}}}{r}\right],
$$
Here we use the dimensionless variables $M=m M_{\odot},\quad
r\rightarrow r_{g}r, \quad \rho\rightarrow\rho
M_{\odot}/r_{g}^{3},\quad P\rightarrow p M_{\odot}c^{2}/r_{g}^{3},
\quad R\rightarrow {R}/r_{g}^{2}$, $\alpha
r_{g}^{2}h(R)\rightarrow \alpha h(R)$, where
$r_{g}=GM_{\odot}/c^{2}=1.47473$ km.

For the Ricci curvature scalar one can get the following equation:
\begin{equation}\label{TOV-3}
3\alpha
r_{g}^{2}\left\{\left[\frac{2}{r}-\frac{3m}{r^{2}}-\frac{dm}{rdr}-\left(1-\frac{2m}{r}\right)\frac{dp}{(\rho+p)dr}\right]\frac{d}{dr}+
\left(1-\frac{2m}{r}\right)\frac{d^{2}}{dr^{2}}\right\}h_{{R}}+\alpha
r_{g}^{2} h_{{R}}{R}-2\alpha r_{g}^{2} h-{R}=-8\pi({\rho}-3p)\,.
\end{equation}
We need to take into account the EoS for matter inside the star
for  the system of  Eqs. (\ref{TOV-1}), (\ref{TOV-2}), (\ref{TOV-3}).

The Lagrangian density for nuclear matter consisting of baryon
octet with masses $m_{b}$ ($b=$$p$, $n$, $\Lambda$,
$\Sigma^{0,\pm}$, $\Xi^{0,-}$) interacting with scalar $\sigma$,
isoscalar-vector $\omega_\mu$ and isovector-vector ${\rho}_{\mu}$
meson fields and leptons ($l=$$e^{-}$, $\mu^{-}$) is
\begin{equation}
\mathcal{L}=\sum_{b}\bar{\psi}_{b}\left[\gamma_{\mu}(i\partial^{\mu}-g_{\omega
b}\omega^{\mu}-\frac{1}{2}g_{\rho
b}{\tau_{3_{b}}}{\rho}^{\mu})-(m_{b}-g_{\sigma
b}\sigma)\right]\psi_{b}+\sum_{l}\bar{\psi}_{l}\left(i\gamma_{\mu}\partial^{\mu}-m_{l}\right)\psi_{l}+
\end{equation}
$$
+\frac{1}{2}\left[(\partial_{\mu}\sigma)^{2}-m^{2}_{\sigma}\sigma^{2}\right]-
V(\sigma)+\frac{1}{2}m^{2}_{\omega}\omega^{2}-\frac{1}{4}\omega_{\mu\nu}\omega^{\mu\nu}-
\frac{1}{4}{\rho}_{\mu\nu}{\rho}^{\mu\nu}+
\frac{1}{2}m^{2}_{\rho}{\rho}_{\mu}{\rho}^{\mu}.
$$
The mesonic field strength tensors are given by relations
$\omega_{\mu\nu}=\partial_{\mu}\omega_{\nu}-\partial_{\nu}\omega_{\mu}$,
$\rho_{\mu\nu}=\partial_{\mu}\rho_{\nu}-\partial_{\nu}\rho_{\mu}$.
The isospin projection is denoted by $\tau_{3_{b}}$. Scalar field
potential $V(\sigma)$ depends on chosen model. The strong
interaction couplings $g_{b\sigma}$, $g_{b\omega}$ and $g_{b\rho}$
depend from density (for details see \cite{Typel}).

Using the mean-field approximation,  one  obtains the following
equations for meson fields:
\begin{equation}\label{0}
m^{2}_{\sigma}\sigma+\frac{dV}{d\sigma}=\sum_{b}g_{\sigma b}
n_{b}^{s},\quad m^{2}_{\omega}\omega_{0}=\sum_{b}g_{\omega b}
n_{b},\quad m^{2}_{\rho}\rho_{0}=\sum_{b}g_{\rho b}
\tau_{3_{b}}n_{b},
\end{equation}
where $\sigma$, $\omega_{0}$, $\rho_{0}=$ are expectation values
of the meson fields. The scalar and vector number densities of
particles are $n^{s}_{b}$ and $n_{b}$ correspondingly. We consider
the GM2 and GM3 parametrization (the nucleon-meson couplings and
parameters of scalar field potential
$V(\sigma)=(1/3)bm_{n}(g_{\sigma n}\sigma)^{3}+(1/4)c(g_{\sigma
n}\sigma)^{4}$ are given in table 1). The hyperon-meson couplings
are assumed to be fixed fractions of nucleon-meson couplings, i.e.
$g_{iH}=x_{iH}g_{iN}$, where $x_{\sigma H}=x_{\rho H}=0.600$,
$x_{\omega H}=0.653$ (see \cite{Rabhi}).

\begin{table}
\label{Table1}
\begin{centering}
\begin{tabular}{|c|c|c|c|c|c|c|c|c|}
  \hline
   & $n_{s}$ & $-B/A$ &  & $g_{\sigma N}/m_{\sigma}$ & $g_{\omega N}/m_{\omega}$ & $g_{\rho N}/m_{\rho}$ &  &  \\
  Model & (fm$^{-3}$) & (MeV) & $M^{*}/M$ & (fm) & (fm) & (fm) & b & c \\
  \hline
  GM2 & 0.153 & 16.30 & 0.78 & 3.025 & 2.195 & 2.189 & 0.003478 & 0.01328 \\
  GM3 & 0.153 & 16.30 & 0.78 & 3.151 & 2.195 & 2.189 & 0.008659 & -0.002421 \\
  \hline
\end{tabular}
\caption{The nucleon-meson couplings and parameters of scalar
field potential for the GM2 and GM3 model \cite{Glendenning}. The
nuclear saturation density $n_{s}$, the Dirac effective mass
$M^{*}$ and the binding energy ($B/A$) are also given.}
\end{centering}
\end{table}

The scalar densities for baryons are given by
\begin{equation}
n^{s}_{b}=\frac{m_{b}^{*2}}{2\pi^2}\left(E_{b}^{f}k^{b}_{f}-
m_{b}^{*2}\ln\left|\frac{k_{b}^{f}+E_{b}^{f}}{m_{b}^{*}}\right|\right),
\end{equation}
where $m_{b}^{*}=m_{b}-g_{\sigma b}\sigma$ is the effective baryon
mass. For the vector densities for baryons we have
\begin{equation}
n_{b}=\frac{1}{3\pi^2} k^{f3}_{b}.
\end{equation}
Here $E^{f}_{b}$ is the Fermi energy, for baryon $E^{f}_{b}$ is
related to the Fermi momentum $k^{f}_{b}$ as $E^{f}_{b}=(k^{f
2}_{b}+m^{*2}_{b})^{1/2}$.

For chemical potential of baryons and leptons, one has
$$
\mu_{b}=E^{f}_{b}+g_{\omega b} \omega_{0}+g_{\rho
b}\tau_{3_{b}}\rho_{0}+\Sigma_{0}^{R},\quad \mu_{l}=E^{f}_{l}.
$$
The rearrangement self-energy term is defined by
\begin{equation}
\Sigma^{R}_{0}=-\frac{\partial \ln g_{\sigma N}}{\partial
n}m^{2}_{\sigma}\sigma^{2}+\frac{\partial \ln g_{\omega
N}}{\partial n}m^{2}_{\omega}\omega^{2}_{0}+\frac{\partial \ln
g_{\rho N}}{\partial n}m^{2}_{\rho}\rho^{2}_{0}.
\end{equation}
Here $n=\sum_{b} n_{b}$. The following conditions should be
imposed on the matter for obtaining the EOS:
\\$(i)$ baryon number
conservation:
\begin{equation}\label{1}
\sum_{b}n_{b}=n,
\end{equation}
$(ii)$ charge neutrality:
\begin{equation}\label{2}
\sum_{i} q_{i}n_{i}=0,\quad i=b,l,
\end{equation}
$(iii)$ beta-equilibrium conditions:
\begin{equation}\label{3}
\mu_{n}=\mu_{\Lambda}=\mu_{\Xi^{0}}=\mu_{\Sigma^{0}}, \quad
\mu_{p}=\mu_{\Sigma^{+}}=\mu_{n}-\mu_{e},\quad
\mu_{\Sigma^{-}}=\mu_{\Xi^{-}}=\mu_{n}+\mu_{e},\quad
\mu_{e}=\mu_{\mu}.
\end{equation}
At given $n$,  Eqs. (\ref{0})-(\ref{3}) can be numerically solved. The resulting EOS are sufficiently soft (for analytical
parametrization see \cite{Eksi}) and in GR,  one
cannot obtain the stars with maximal mass $\sim 2M_{\odot}$ (see
Fig. 1, curves labelled as gm2nph and gm3nph). However, as we
demonstrate below, the situation is qualitatively different in
modified gravity.

For the solution of Eqs. (\ref{TOV-1})-(\ref{TOV-3}),  one can use a
perturbative approach (see \cite{Arapoglu,capo,Alavirad} for
details). For a perturbative solution, the density, pressure, mass
and curvature can be  expanded as
\begin{equation}\label{expan}
p=p^{(0)}+\alpha p^{(1)}+...,\quad \rho=\rho^{(0)}+\alpha
\rho^{(1)}+...,
\end{equation}
$$
m=m^{(0)}+\alpha m^{(1)}+...,\quad R=R^{(0)}+\alpha R^{(1)}+...,
$$
where functions $\rho^{(0)}$, $p^{(0)}$, $m^{(0)}$ and $R^{(0)}$
satisfy the standard TOV equations. Terms containing $h(R)$ are
assumed to be of first order in the small parameter $\alpha$, so
all such terms should be evaluated at ${\mathcal O}(\alpha)$
order.

Finally, perturbative TOV equations are, for mass $m=m^{(0)}+\alpha
m^{(1)}$:
\begin{equation}\label{PTOV-1}
\frac{dm}{dr}=4\pi\rho r^2-\alpha r^{2}\left[4\pi
\rho^{(0)}h^{(0)}_{R}+\frac{1}{4}\left(h^{(0)}-h^{(0)}_{R}R^{(0)}\right)\right]+\frac{1}{2}\alpha\left[\left(2r-3m^{(0)}-
4\pi\rho^{(0)}r^{3}\right)\frac{d}{dr}+r(r-2m^{(0)})\frac{d^{2}}{dr^{2}}\right]
h^{(0)}_{R}
\end{equation}
and for pressure $p=p^{(0)}+\alpha p^{(1)}$:
\begin{equation}
-\left(\frac{r-2m}{\rho+p}\right)\frac{dp}{dr}=4\pi r^2
p+\frac{m}{r}-\alpha r^2\left[4\pi
p^{(0)}h^{(0)}_{R}-\frac{1}{4}\left(h^{(0)}-h^{(0)}_{R}R^{(0)}\right)\right]-
\alpha \left(r-\frac{3m^{(0)}}{2}+2\pi
p^{(0)}r^{3}\right)\frac{dh^{(0)}_{R}}{dr}.
\end{equation}

 In
\cite{Astashenok}, the Mass-Radius relation for the neutron stars,
in particularly, for modified gravity with $f(R)=R+\alpha
R^2(1+\gamma R)$ is considered. In that case, we found that, for
high central densities  a  second ``branch'' of stability emerges
with respect to the one existing in GR. This
stabilization of star configurations occurs due to the presence of cubic term in the Ricci curvature scalar.

\begin{figure}
  \includegraphics[scale=1.1]{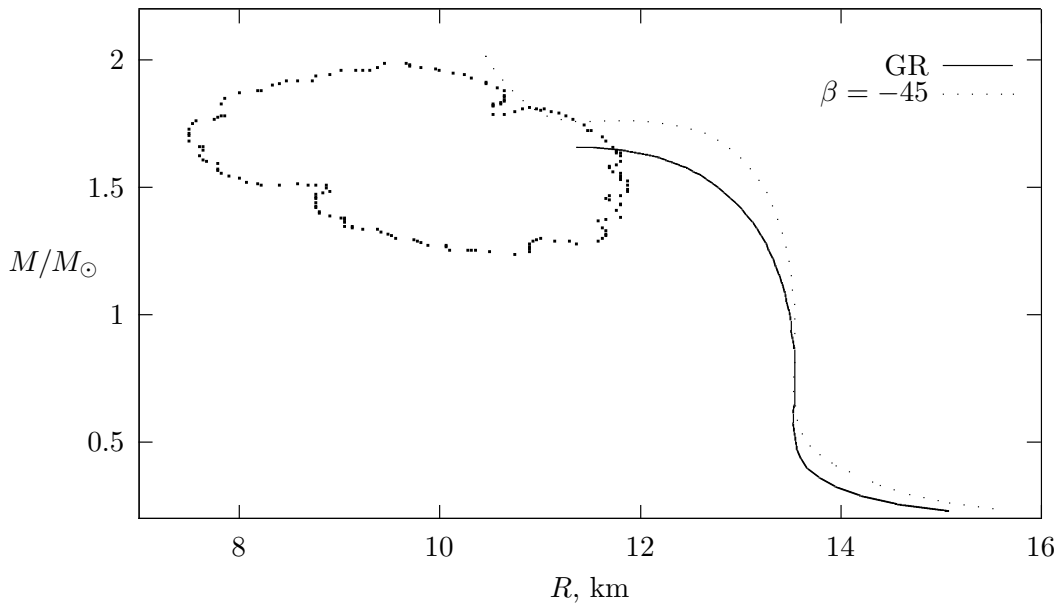}\\
  \caption{The Mass-Radius diagram for GM2 model extended to hyperon
  sector (gm2nph) in modified gravity model $f(R)=R+\beta R^3$ and in GR, for comparison. For $\beta\approx -45$ (in units of $r_{g}^{4}$), the maximal limit of
  mass for star is around $2M_{\odot}$. The corresponding central density is $3.48\times 10^{15}$ g/cm$^{3}$.}
\end{figure}

\begin{figure}
  \includegraphics[scale=1.1]{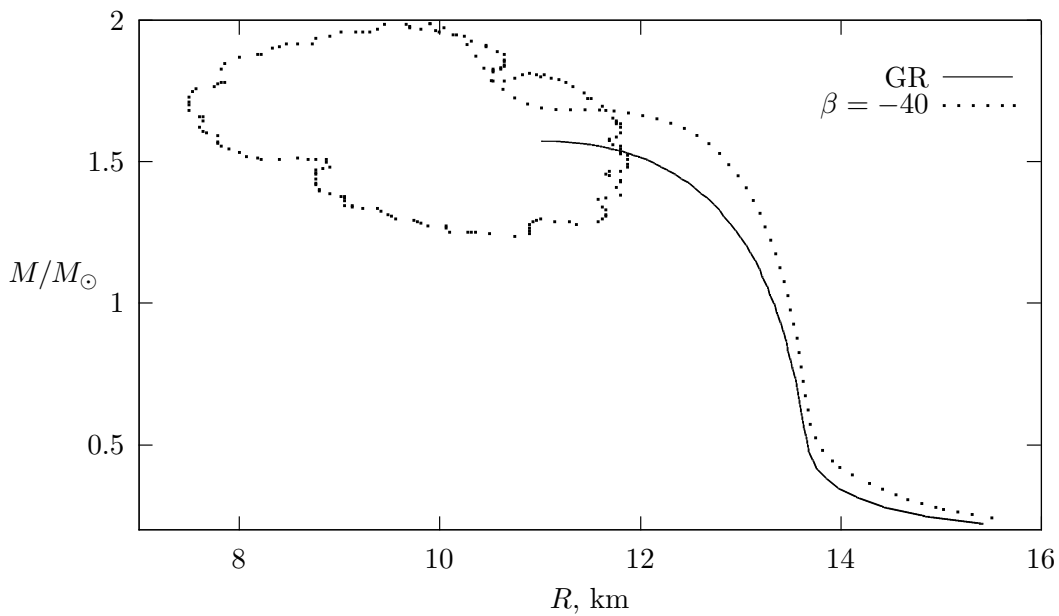}\\
  \caption{The Mass-Radius diagram for gm3nph model in modified gravity $f(R)=R+\beta R^3$ and in GR for comparison. For $\beta\approx -40$ (in units of $r_{g}^{4}$) the maximal limit of
  mass for star is around $2M_{\odot}$. The corresponding central density is $3.34\times 10^{15}$ g/cm$^{3}$.}
\end{figure}

For modified gravity with only cubic term (that is $f(R)=R+\beta R^{3}$)
the maximal value of neutron star mass for given EOS increases for
$\beta<0$. This effect allows to construct  neutron star models
with maximal mass $\sim2M_{\odot}$ even for those hyperon EOS
which do not satisfy  the observational constraints coming from  standard GR. In other words, these stable star configurations can exist at higher
central densities than in GR.

In  Figs. 2 and 3,  the Mass-Radius diagram for simple hyperon models
(gm2nph and gm3nph) with realistic parameters is represented. We
define the values of the parameter $\beta$ for obtaining the star
configurations with $M\sim 2M_{\odot}$.

Note that the dimensionless parameter $\alpha$ in modified TOV
equations can be defined in our case as $\alpha=\beta
|R^{(0)3}|_{max}$, where ``max'' means maximal value of cubic term
at ${\mathcal O}(\alpha)$ order. The scalar curvature $R^{(0)}$ is
simply
$$
R^{(0)}=8\pi (\rho^{(0)}-3p^{(0)}).
$$
One can determine the dimensionless parameter
$$
\delta=\beta R^{(0)2}.
$$
In Fig. 4,  the dependence of this parameter from density (for star
configuration with maximal mass $\sim 2 M_{\odot}$) is represented
for gm2nph and gm3nph model. One can see that the cubic term is small if
compared to $R$ even for high central densities.

\begin{figure}
\includegraphics[scale=1.0]{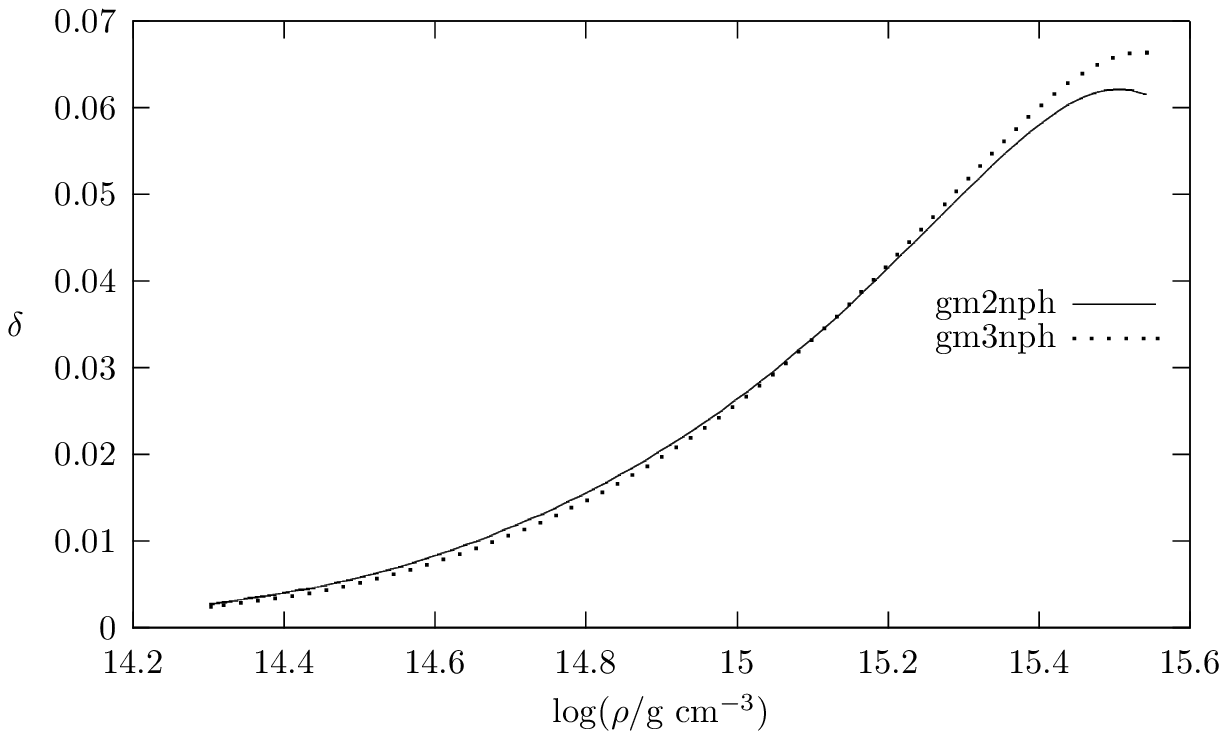}\\
  \caption{The density dependence of dimensionless parameter $\delta=\beta
  R_{0}^2$ for gm2nph ($\beta=-40$) and gm3nph ($\beta=-45$) models in cubic gravity.
The maximal value is less than $0.1$ even for central regions of
star.}
\end{figure}

\begin{figure}
\includegraphics[scale=1.0]{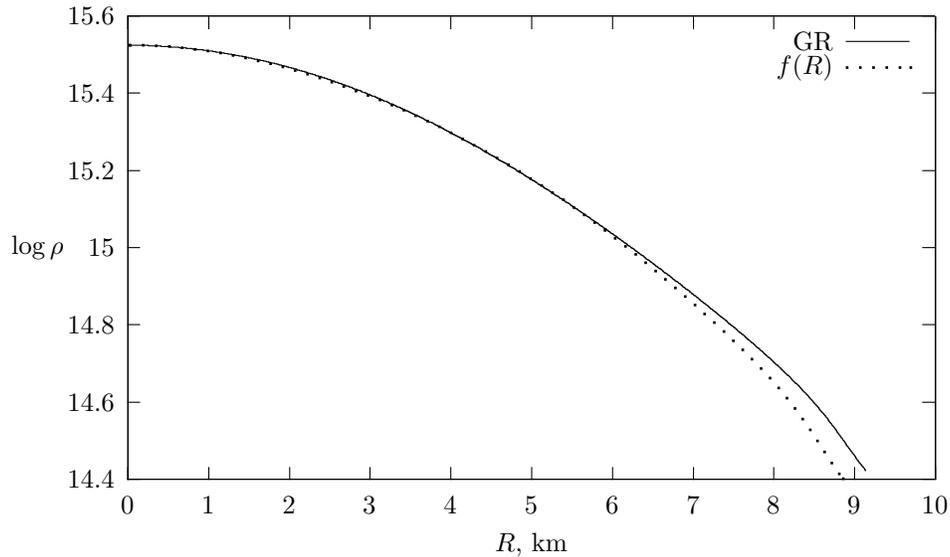}\\
  \caption{The density profile for gm3nph model in modified gravity (for $\beta=-40$ and $\rho_c=3.34\times 10^{15}$ g/cm$^{3}$) in comparison with the one  in GR.
  The difference is insignificant at high densities. The increase of ``effective'' ($\rho_{eff}=\frac{1}{4\pi^2}\frac{dm}{dr}$) density (and maximal mass)  occurs due to the terms containing $\alpha$ in r.h.s. of Eq.(\ref{PTOV-1}). A similar picture takes place for gm2nph model.}
\end{figure}

The density profile even for star configuration with maximal mass
$2M_{\odot}$ is almost the same as the corresponding profile for
star model in GR (see Fig. 5). The increase of the ``effective''
density (and maximal mass)  occurs due to the terms containing
$\alpha$ in r.h.s. of  Eq. (\ref{PTOV-1}).

The increase of maximal neutron star mass occurs for
 realistic $f(R)$ model of gravity with quadratic and cubic terms, that is
\begin{equation}
f(R)=R+\gamma R^{2}+\beta R^{3}.
\end{equation}

The effect occurs for $\gamma<0$ if cubic term is greater than
quadratic at high densities. For a given value of $\gamma$, one can
define the parameter $\beta$ where the maximal value of neutron
star mass is $\sim 2 M_{\odot}$.

\begin{figure}
  \includegraphics[scale=1.1]{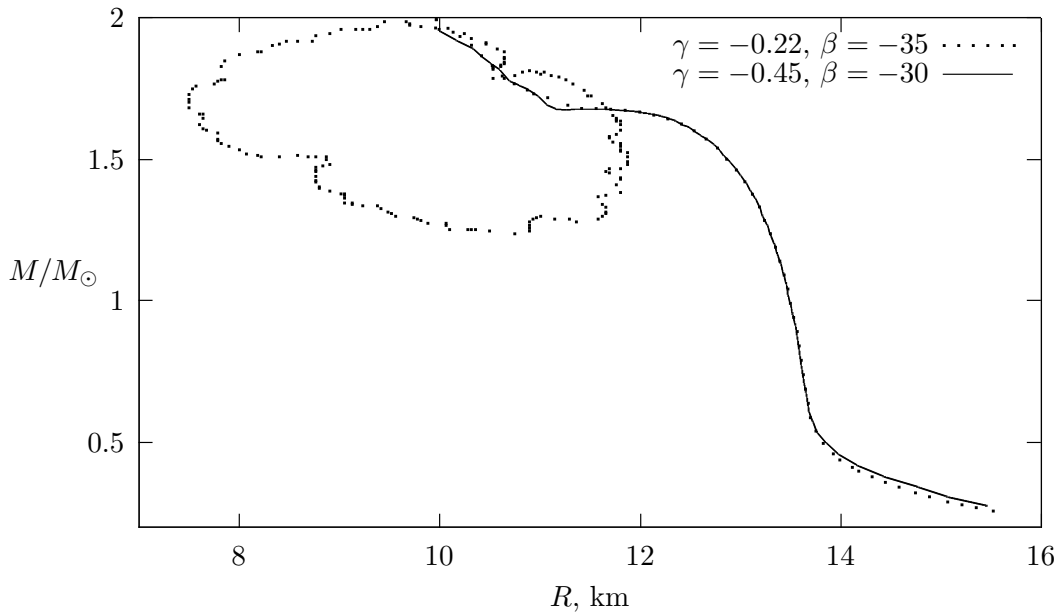}\\
  \caption{The Mass-Radius diagram for gm3nph EOS in modified gravity
$f(R)=R+\gamma R^{2}+\beta R^3$ with maximal mass $\sim 2M_{\odot}$ for two values of $\gamma$. These curves are close
  to M-R relation in the model with only the  cubic term (see Fig. 3). Note that the quadratic term is
  smaller than the cubic one  for given EOS if $\rho>\rho_{0}\approx 2\times 10^{14}$ g/cm$^{3}$.}
\end{figure}

Furthermore, let us  consider the case of gm3nph EOS. In Fig. 6,
the realistic Mass-Radius diagram is represented for two values of
$\gamma=5\times 10^{9}$ cm$^{2}$, $1\times 10^{10}$ cm$^{2}$ (or
$\sim 0.22$ and $\sim 0.45$ in units of $r_{g}^{2}$). One can see
that the two solar mass limit is reached for $\beta\approx -35$
and $\beta\approx -30$ for these values of $\gamma$. In fact, the
$M-R$ relation,  in this case, is close to the $M-R$ relation for
$f(R)$ model without quadratic term. The analysis shows that there
is a set of parameters ($\gamma$; $\beta$) at which we have the
same $M-R$ relation. Hence, analytical $f(R)$ gravity models  with
quadratic and cubic terms may provide the resolution of neutron
star maximal mass and hyperonization puzzle problems,  being
consistent, at the same time,  with the M-R diagram.

\section{Conclusions and perspectives}

In summary, we presented a possible solution of the ``hyperonization puzzle''
in the neutron star theory. The softening of nucleon EOS, due to
hyperonization, leads to the decrease of the upper limit mass of neutron
star considerably below the  two solar masses (in the simple model
of hyperonic matter with realistic parameters) according to GR. However, in
modified $f(R)$ gravity model with cubic  and quadratic terms,  it is
possible to obtain neutron  stars with $M\sim 2M_{\odot}$ for simple
EOS from GM2 and GM3 model extended to hyperon sector. Of course,
modified gravity under consideration is chosen to be in power-law analytic
form as a simple example. However, the preliminary estimations
indicate that similar effect may be expected for viable modified
gravities where the  analysis is more detailed and realistic stellar models are considered.
Note also that power-law $f(R)$ models  are the  standard approximation for
more complicated non-linear $f(R)$ gravities. However, it is important to point out that the Mass-Radius
relation significantly differs from GR  only at
high central densities. As  consequence, the ``effective'' EOS
is sufficiently soft to describe the radii and masses measurements
for the  three observed neutron stars EXO 1745-248, 4U 1608-52 and  4U
1820-30. In other words, the same modified gravity may solve
simultaneously three problems of neutron stellar astrophysics:
maximal mass of neutron star, realistic Mass-Radius relation and
hyperonization puzzle.

However, some final important   remarks are useful at this point.
The hyperonization puzzle can also be solved by considering alternatives to  the $SU(6)$  model
 which prescribes particular relations between the hyperon-meson and  nucleon-meson
 couplings. Because the nature and coupling of the light sigma meson is not  known,
 it is most natural to look for improving  this meson model  (see \cite{armen1}).  Another
 solution is that the hyperonic matter appears only at intermediate
 densities and the  formation of quark cores supports hypernuclear stars agains collapse (see
\cite{armen2,armen3}. It should be clear that there is enough room in the
 nuclear theory which,  combining  conventional gravity, can allow for massive
 stars. Of course this does not exclude alternative gravity as a possible solution   of the problem.

As a  next step, we will
extend our results to non-perturbative treatment of TOV
equations. However, up to now,  it seems a   very hard problem (see, for
instance,\cite{rad}) which may need the development of
qualitatively new numerical methods due to higher-derivative structure of
$f(R)$ gravity and necessity to account for {\it chameleon effects} as
well as quantum gravity effects at very high densities.

\acknowledgments
{This work is supported in part by projects 14-02-31100 (RFBR,
Russia) and 2.2529.2011 (MES, Russia) (AVA), by JSPS Short-Term
Program S-13131 (Japan), by MINECO (Spain), FIS2010-15640 and by
MES project 2.1839.2011 (Russia) (SDO). SC is supported by INFN ({\it iniziative specifiche} TEONGRAV and QGSKY).}

\end{document}